# Maximal Density, Kinetics of Deposition and Percolation Threshold of Loose Packed Lattices

I. Avramov, V. Tonchev

R. Kaischew Institute of Physical Chemistry, Bulgarian Academy of Sciences,

1113 Sofia, Bulgaria

**Abstract**

In many areas of research it is interesting how lattices can be filled with particles that have no nearest neighbors, or they are in limited quantities. Examples may be found in statistical physics, chemistry, materials science, discrete mathematics, etc. Using Monte Carlo (MC) simulation we study the kinetics of filling of square lattice (2D). Two complementary rules are used to fill the lattice. We study their influence on the kinetics of the process as well as on the properties of the obtained systems. According to the first rule the occupied sites may not share edges (nearest neighbors occupations are not permitted). Under this condition, the maximum possible concentration is 0.5, forming a checkerboard type structure. However, we found that if the deposition is done by random selection of sites the concentration of 0.5 is inaccessible and the maximum concentration is $C_{max}(2D)=0.3638 \pm 0.0003$ for 2D lattice. If the lattice is 3D we find that the maximal concentration is even lower $C_{max}(3D)=0.326 \pm 0.001$. The second rule establishes permission to break the first one with certain probability $0 \leq p \leq 1$, thus the occupied sites can start to share edges when $p>0$. In this case higher then *0.3638* concentrations are accessible, even up to *C=1*. Therefore the percolation threshold $P_c$ can be reached. Its value depends on the value of the probability *p*. Our model describes the kinetics of formation of thin films of particles attracted by the substrate but repulsing each other.

**Introduction**

Percolation theory was developed to describe flow in porous media [1]. Later, this model has attracted the interest of scientists working in different fields [2, 3]. The reason is that the concentration of the percolation threshold, $P_c$, is important not only for the extraction of oil from porous minerals. In materials science, percolation theory addresses the problem why some important properties are well explained by assuming that a network of molecules creates a rigid system and other properties of the same materials behave as if the network is soft. For example, solid electrolyte materials have a rigid carcass while at the same time a set of flexible channels permeates through. Rigid and flexible networks are discussed in detail in [4-8]. Rigidity percolation of clusters at the glass transition temperature is used to explain the structural details of a class of metallic glasses [9]. Percolation in networks (or lattices) offers realistic description of many problems like the nature of social contacts, the spread of diseases, etc. The dynamics of opinion sharing and competing [10] is a particular case of percolation problems.

In most of the cases the value of $P_c$ cannot be determined analytically, so it is estimated numerically, even when the cluster definition is rather complex [11]. A tempting possibility appears to vary the exact location of the critical point by appropriately handling the rules by which the system is built up [12-19]. New models appeared [12, 17] that vary systematically the way the lattice is filled up. As a result different critical threshold values were obtained.

Frequently, particles at interfaces tend to repulse each other, still if the energy gain from the attachment to the interface exceeds this repulsion, they may form a layer. Here interface may be either the boundary between two liquids or the surface of foreign solid substrate. The kinetics of formation of this layer is subject of the present study. Note that if the repulsive energy is large enough, rule arises that prohibits particles, deposited on the interface, to have neighbours in their immediate environment. So, upper limit of concentration appears which cannot be exceeded. However, if there is certain probability *p>0* to neglect this rule, the kinetics of deposition undergoes a crossover in the vicinity of the *jammed* concentration and continues at much lower rate. The value of *0 ≤ p ≤ 1* is a measure of the balance between the attractive and repulsive tendencies.

## 1. The Model

We consider deposition processes on square lattice of size *L*. During deposition the sites transform irreversibly from vacant state to occupied state. The act of such transformation is referred to as particle deposition. The overall number of sites is $L^2$. In the present study we use a two-dimensional lattice to provide a ground for building the conceptual model as simple as possible and then to increase further its complexity. The concentration *C* is the ratio of the occupied sites to the total number of sites $L^2$. The fraction of the unoccupied sites is *1-C*. At each successive step we increase the number of occupied sites with one on the expense of the number of unoccupied.

We study the percolation threshold $P_c$. By Kolmogorov's zero-one law, for any given $C$, the probability that an infinite cluster exists is either zero or one [20]. Since this probability is an increasing function of $C$, there must be a critical value, $P_c$, below which the probability is 0 and above which the probability is 1. The probability of an open path from a side (of the finite lattice) to the opposite one increases sharply from very close to zero to very close to unity in a short span of values of $C$. As percolation can be formulated as the emergence of long-range connectivity in large random systems [21-25], in addition to $P_c$, we also study the concentration $P_p$ at which the largest cluster touches, for the first time, two opposite sides of the lattice. We find that for all lattices, having size $L>500$, the values of $P_c$ and $P_p$ are equal, $P_c \approx P_p \pm 0.004$.

In our model, the deposition is regulated by two rules:

(1) if the chosen site has no neighbors, then it is occupied;

(2) if the chosen site has occupied neighbors, then it is occupied with probability $0 \leq p \leq 1$.

If $p=1$, occupation (deposition) takes place, no matter whether some of the neighbors is occupied (i.e. the sites sharing edges are permitted explicitly). In the other limiting case, $p=0$, the deposition on sites with already occupied neighbours is forbidden.

We measure the time in MC steps. To make time independent on the lattice size, the relative time, $t$, is the number of MC steps divided by the overall number $L^2$ of lattice sites, i.e. the relative time, $t$, is the average number of MC steps per site. If $p>0$, clusters can be formed when occupying sites that share edges. We measure the concentration $C$ of the occupied sites and the dimensionless percolation strength, $0<S_m<1$, which gives the probability of a given occupied site to belong to the largest cluster. If the number of occupied sites belonging to the largest cluster is $N_m$ then $S_m$ is determined as:

$$S_m = \frac{N_m}{C.L^2} \qquad (1)$$

Traditionally, the percolation threshold $P_c$ is determined as the inflection point [12-19] of the curve of $S_m$ against $C$. Periodic boundary conditions are used, except for $P_p$ determination.

## 2. Results and Discussion

The present model is suitable to treat kinetics of formation of two-dimensional interfaces of repulsing particles. In materials science, examples of such systems are protein solutions and gas phases. Indeed, the molecules of the solvent in solutions, as well as in gases, tend to keep separated. When concentration of the solution (respectively the pressure in the gas phase) increases we reach the jamming concentration. This is break point for systems with $p=0$. If particles are repulsing strongly each other further increase of the concentration is impossible. Anticipating, we note that for $p>0$, i.e. the repulsion force is not so strong, further deposition remains possible, although at much slower rate.

### a. Random deposition ( p=1)

The condition $p=1$ corresponds to random deposition because the rule about the number of occupied neighbours is supressed. As this case is well studied [2,3,16], here it is used to validate our code. The time dependence of $C$ is linear with a slope equal to one, see Fig.1. Results for all lattice sizes with $L \geq 100$ are overlapping.

The dependence of $S_m$ on $C$ is also studied. Despite that the critical threshold increases continuously with $L$, above $L>500$ the increase is not essential, so that $P_c \approx 0.5927$ is sufficiently accurate for $p=1$.

### b. Unlimited repulsion of nearest neighbours: ( p=0).

Under this condition the particles do not contact, or at least prefer to be in contact with solvent molecules. Clearly, percolation is not possible because the occupied sites cannot share edges, i.e. particles do not touch. It can be shown that the densest packing is checkerboard like lattice with maximum concentration $C_{max} = 0.5$. This structure itself is subject to studies [26].

On large lattices the random deposition leads to the formation of tiny regions with checkerboard like structures separated by channels on which deposition is forbidden. This is demonstrated in Fig.2, where a cross section 20x20 from a lattice with $L=1000$ is shown. Note that the maximal concentration $C_{max}$ that can be achieved is less than 0.5. In a sense, the maximal concentration is similar to what in RSA studies [24,25] is called jammed concentration. Our MC simulation has shown that although $C_{max}$ depends on the lattice size $L$, for large lattices, if deposition is random, than there is a limiting value of $C_{max}=0.3638 \pm 0.0003$, as shown in Figs.(1 and 3). This value is close to predictions [24, 25] for other systems. The "dip" around $L=500$ in the $C_{max}$ vs. $L$ curve (Fig.3) rather suspicious. At present stage the accuracy with which the value of $C_{max}$ was determined is about $10^{-4}$. We briefly studied this problem for 3D lattice and find that in this case the maximal concentration is even lower, $C_{max}(3D)=0.326 \pm 0.001$.

### c. Limited repulsion of nearest neighbours: ( $p>0$ ).

The situation changes completely if $p>0$. This condition corresponds to comparable contributions of two energies: the energy $E_s$ of attraction of the particle to the interface and the repulsion energy $\sum_{i=1}^{4} E_i$ between the neighbours.

In materials science $p$ is frequently the Boltzmann probability and can be given as:

$$p = \begin{cases} 1 & , E<0 \\ e^{-\frac{E}{k_B T}} & , E \geq 0 \end{cases} \qquad (2)$$

Here energy $E$ stands for $E = E_s - \sum_{i=1}^{4} E_i$. Note that Eq.(2) is just an example of one of the possible models to determine $p$. Detailed models on $p$ (for instance verification of Eq.2) are task for further treatments. In this article we just give some numerical values for $p$ as examples. The decrease of $p$ causes the molecules to separate. Increasing the fraction $C$ of occupied sites corresponds to increase of pressure in gas systems.

Fig.1 presents the kinetics of formation of thin films, i.e. the time dependence of the concentration C of occupied sites. It is given for several values of p. We find that the time dependence is independent on the lattice size $L$ in the interval $100<L<10\,000$. For a given $p>0$ value the results from all lattices, having size $L>100$, lay on the same master curve, so that results shown in Fig.1 are valid for any $L>100$. The time (the average number of MC steps per lattice site) for which concentration of occupied sites approaches $C=1$ is about $1/p$. The data are for several values of the probability $p$ as follows: The thick dashed line is for $p=0$. It gives the limiting condition $C_{max} = 0.3638$ of maximal possible concentration for random deposition. The dotted line is for $p=0.001$; the dashed line is for $p=0.01$ and the dot-dashed line is for $p=0.1$.

Similar is the result for the dependence of the percolation strength $S_m$ on concentration $C$, only this time lattice has to be a little bit larger, $L>500$. The dependencies of percolation strength $S_m$ on concentration $C$ are sigmoidal curves shown in Fig.*4A*. The left side solid line is for *p=1*. The curves are moving to the right as the value of exception probability *p* decreases. The curves illustrating this are for: *p=0.5; p=0.25; p=0.1* and *p=0.001*. The corresponding derivatives are shown in Fig.*4B*.

Fig.5 demonstrates the dependence of percolation threshold $P_c$ on exception probability *p*. The value *p=1* corresponds to random deposition, independent on the number of occupied neighbours. As expected, at *p=1* the percolation threshold is $P_c=0.5927$. The lowest *p* value at which we determined the percolation threshold is 0.001. The value of percolation threshold $P_c$ decreases as *p* increases. This finding is in agreement with the result published recently in [17]. Theoretical treatment of this effect is tempting problem for future studies. So far, the line connecting points is eye guiding. The reason is that at the value *p=0* percolation is impossible because in this case we find that at random deposition one cannot reach concentrations higher $C_{max}=0.3638 \pm 0.0003$.

It is remarkable to note that, for small *p* values, a crossover appears in the dependence of concentration *C* on time *t*. This happens in the vicinity of $C_{max}$ because the occupied sites are now permitted to have occupied nearest neighbours, thus they can share edges, in addition to sharing vertices. Actually, the concentration region at which the process starts to be significant is shortly below $C_{max}$, where the mentioned crossover appears. In this region a sharp decrease of deposition rate appears. Above $C_{max}$ the deposition rate $\frac{dC}{dt} \sim p$ becomes proportional to probability *p* to break the first rule. The crossover corresponds to smooth transition from gaseous to liquid state beyond the tripple point (note that increasing concentration corresponds to increase of pressure in gases). The transition interval is between $C_{max}$ and the percolation threshold $P_c$.

**Conclusions**

The properties, as well as the deposition kinetics, of thin films of particles, repulsing each other, are studied. There is maximal concentration $C_{max}$ that can be reached if occupied sites that are deposited at random can stick with vertices but not with edges. This jamming concentration is $C_{max}=0.3638 \pm 0.0003$ for 2D lattice, and $C_{max}(3D)=0.326 \pm 0.001$ for 3D lattice.

If there is probability $p>0$ to overcome the repulsion from the nearest neighbours, the deposition continues beyond $C_{max}$, although at much lower rate, proportional to $p$. The deposition can go above the percolation threshold $P_c$ but latter depends on the value of $p$ - the order breaking rule.

**Acknowledgments**

This work is done within the frame of the bilateral collaboration between Bulgarian Academy of Sciences and Aristotle University (Thessaloniki) and authors express their gratitude to P. Argyrakis for the stimulating discussions. VT acknowledges the financial support of Bulgarian National Science Fund through grant No. T02-8/121214.

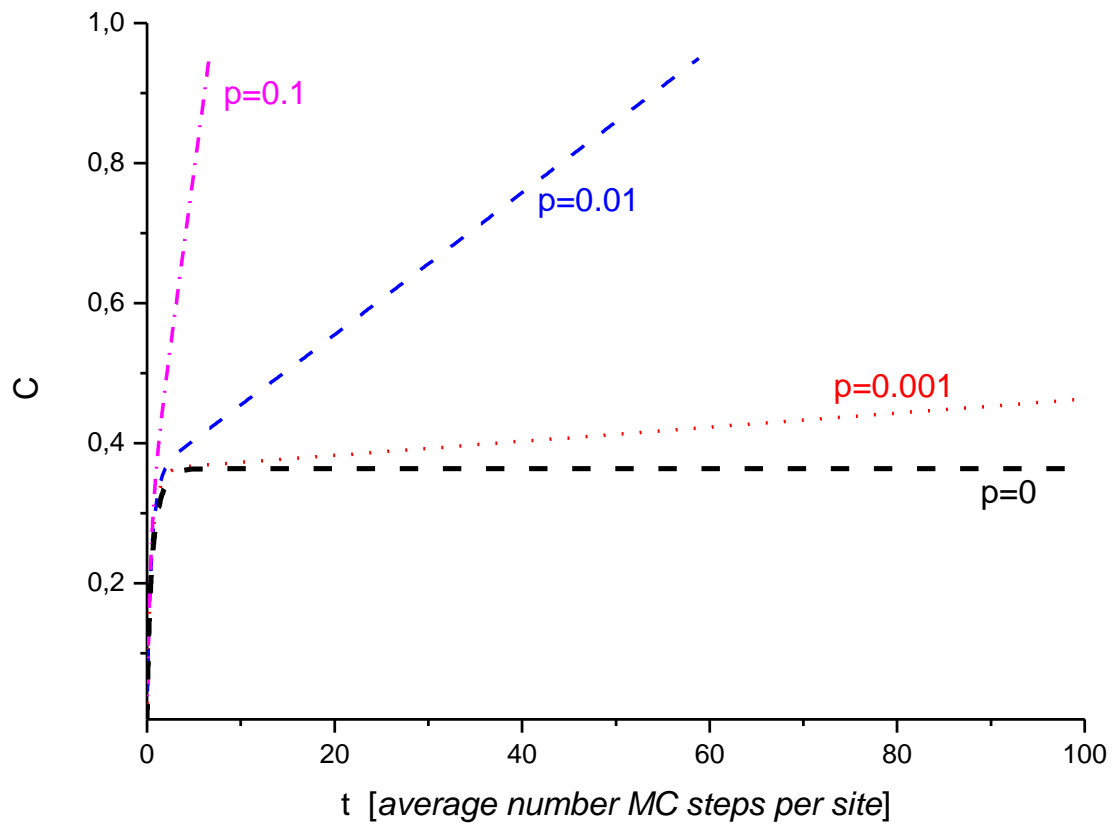

Fig.1

*Fig.1 The time, t, dependence of the concentration C of occupied sites of 2D lattice for several values of the probability, p. Results for all lattice sizes with L≥100 are overlapping. The thick dashed line is for p=0. It gives the limiting condition $C_{max}=0.3638$ of maximal possible concentration for random deposition; the dotted line is for p=0.001; the dashed line is for p=0.01 and the dot-dashed line is for p=0.1.*

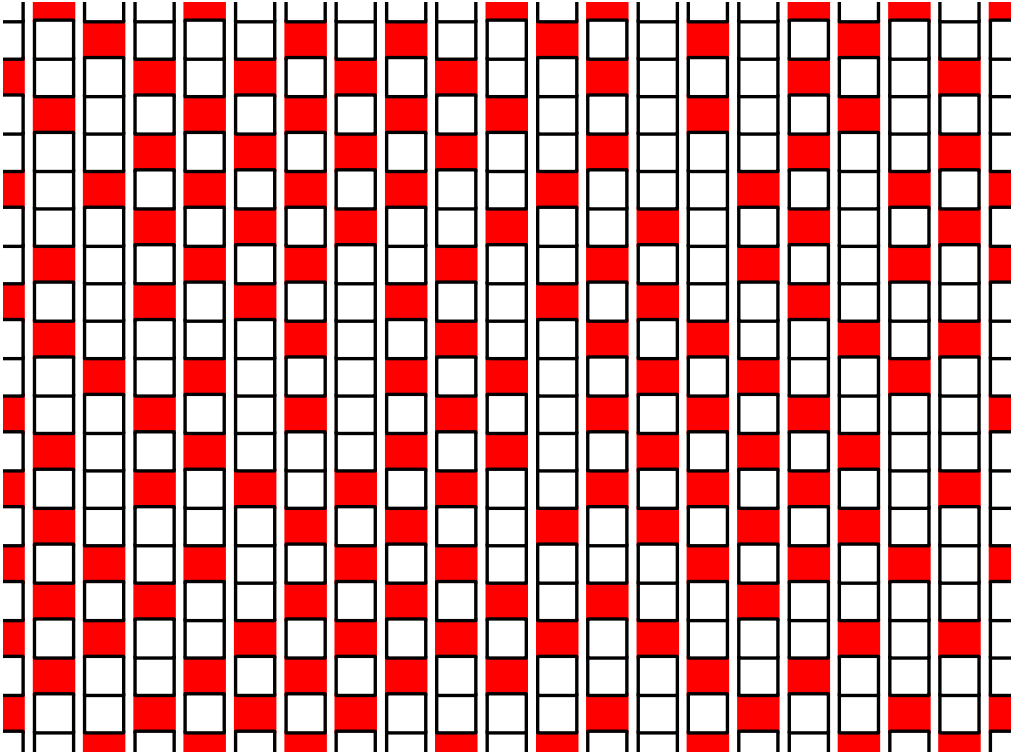

*Fig.2*

*Fig.2 The picture is a cross-section (20x20) of a large (1000x1000) lattice. Occupied sites are marked solid. The concentration is C=0.36*

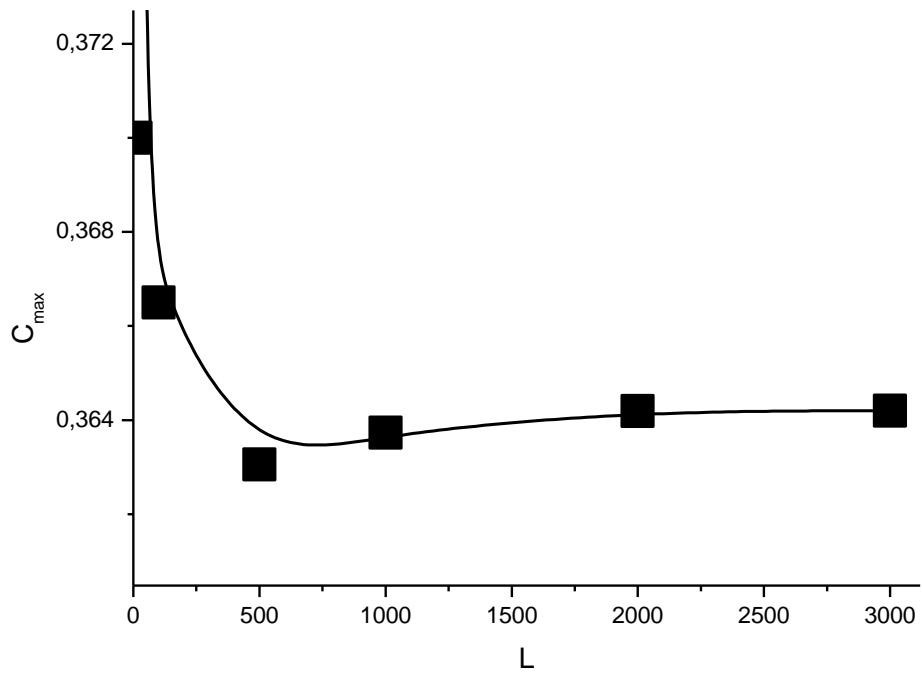

*Fig.3*

*Fig.3 The maximal possible concentration in dependence on the lattice size. The solid line is eye guiding. Random deposition with p=0.*

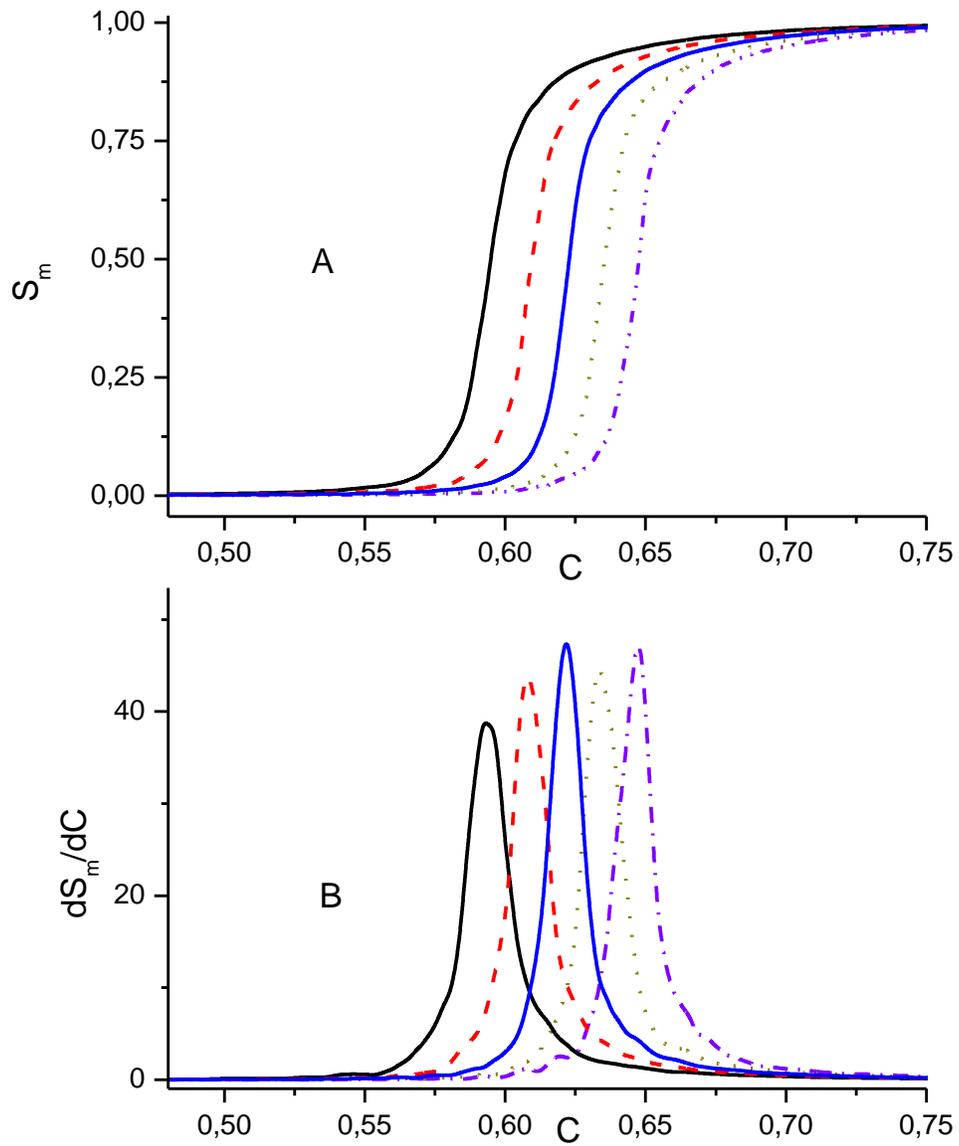

Fig.4

*Fig.4 The dependence of percolation strength $S_m$ and of corresponding derivatives $dS_m/dC$ on concentration C. The lattice size is L=1000; The exception probabilities are (from left to right) as follows: p=1; p=0.5; p=0.25; p=0.1; p=0.001.*

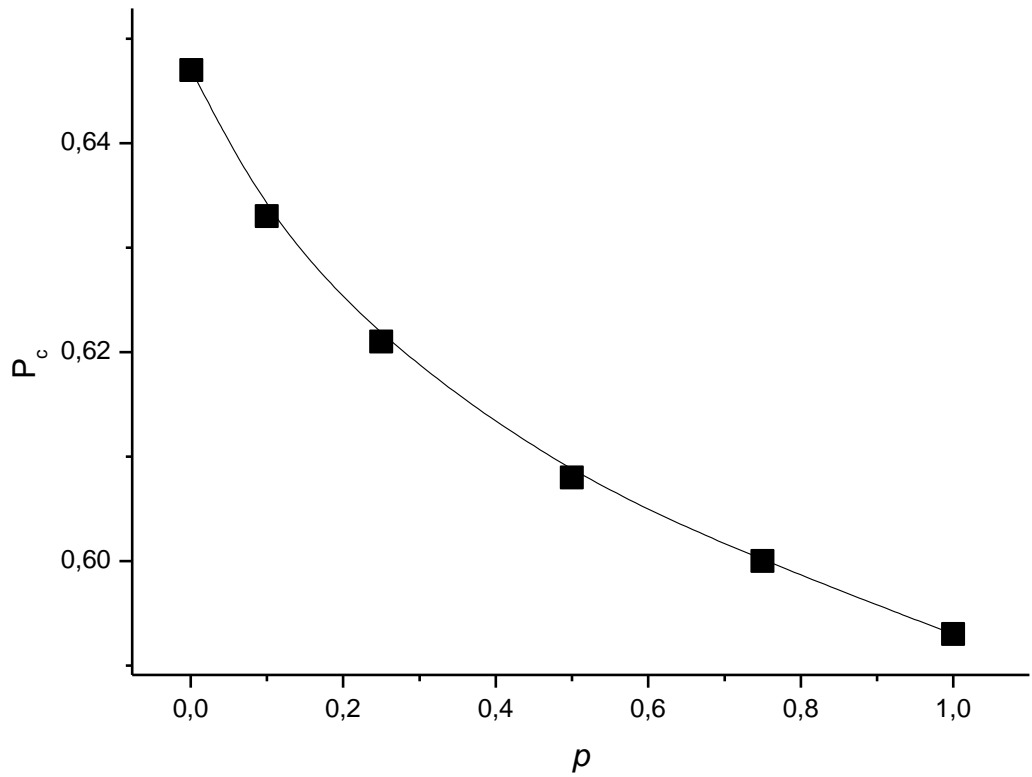

*Fig.5*

*Fig.5 The dependence of percolation threshold $P_c$ on exception probability p. The value p=1 corresponds to random deposition independent on the number of neighbours. The lowest point is for p=0.001. The value p=0 is impossible because in this case there is no percolation.*